\documentclass[12pt]{article}
\usepackage{amsmath}
\usepackage{amssymb}
\usepackage{graphics}

\makeatletter\@addtoreset{equation}{section}
\makeatother

\def\pa{\partial}
\usepackage[dvips]{color}

\begin{document}
\begin{titlepage}
\begin{flushright}
TIT/HEP-611\\
March 2011
\end{flushright}
\vspace{0.5cm}
\begin{center}
{\Large \bf
BPS Monopole Equation in $\Omega$-background}
\lineskip .75em
\vskip1.0cm
{\large Katsushi Ito, Satoshi Kamoshita and Shin Sasaki }
\vskip 2.5em
{\normalsize\it Department of Physics\\
Tokyo Institute of Technology\\
Tokyo, 152-8551, Japan} 
\vskip 3.5em
\end{center}
\begin{abstract}

We study deformed supersymmetries in $\mathcal{N} = 2$ super Yang-Mills theory
in the $\Omega$-backgrounds characterized by two
complex parameters $\epsilon_1,\epsilon_2$.
When one of the $\epsilon$-parameters vanishes, the theory has extended supersymmetries. 
We compute the central charge
of the algebra
and obtain the deformed BPS monopole equation.
We examine supersymmetries preserved by the equation.
\end{abstract}
\end{titlepage}

\baselineskip=0.7cm
%\tableofcontents

%%%%%%%%%%%%%Section 1%%%%%%%%%%%%%%

\section{Introduction} 

The $\Omega$-background deformation of $\mathcal{N} = 2$
supersymmetric gauge theories 
is a useful method to regularize the integrals over the moduli
space of instantons \cite{Moore:1997dj,Nekrasov:2002qd,Losev:2003py}.
This background is characterized by
two anti-symmetric matrices $\Omega^{mn}$ and
$\bar{\Omega}^{mn}$ parametrized  by two complex numbers
$\epsilon_1$, $\epsilon_2$ and their complex conjugates.
The $\Omega$-background
induces the $U(1)^2$ vector fields on ${\bf R}^4$.
This torus action is used to define the supercharge, which is
shown to be equivariantly nilpotent by introducing the Wilson line 
gauge fields.
Using the localization theorem \cite{Nekrasov:2002qd}, 
the regularized 
integral (the instanton partition function) 
leads to the Seiberg-Witten (SW)
prepotential \cite{Seiberg:1994rs} 
in the limit $\epsilon_1,\epsilon_2\rightarrow 0$. 
In superstring theory the $\Omega$-background is realized as a certain
$\mathcal{N}=2$ supergravity background 
\cite{Billo:2006jm,Antoniadis:2010iq,Nakayama:2010ff,IKV,Ta,AwKa,HuKl}.

Recently it has been pointed out that there is a relation
between two-dimensional
integrable systems and gauge theories in the $\Omega$-background where one of the
$\epsilon$-parameters  vanishes
\cite{Nekrasov:2009rc,Nekrasov:2010ka,Maruyoshi:2010iu,
Orlando:2010uu,Poghossian:2010pn}.
The theory in this $\Omega$-background 
has two dimensional $\mathcal{N}=2$ super-Poincar\'e 
invariance.
In particular it was shown
that the instanton partition function in the limit
$\epsilon_2\rightarrow 0$ with $\epsilon_1=\hbar$ is obtained by
the deformation of the SW theory
\cite{Mironov:2009uv,Mironov:2009dv,Mironov:2009ib,Popolitov:2010bz}.
The period integrals of the SW differential are obtained by solving the
quantized Toda spectral curve equations, which 
are evaluated by the exact Bohr-Sommerfeld integrals.

The purpose of this paper is to study the deformed SW theory from a
field theoretical point of view.
The sum of the period integrals of the SW differential is the central charge of 
$\mathcal{N}=2$ supersymmetry algebra \cite{Witten:1978mh}.
We will study the deformed supersymmetry algebra in the
$\Omega$-background and calculate the central charge.
Such deformations of the central charge
have been known
for non(anti)-commutative field theories \cite{Chu:2005me,Ito:2005xe}.
The $\Omega$-background deformation of the central charge
would give $\hbar$-corrections to the BPS spectrum, with which
one can compare the $\Omega$-deformation to the period integral.
In this paper we will derive the deformed BPS monopole equation
in the $\Omega$-background.

The organization of this paper is as follows.
In section 2, 
we introduce the four-dimensional $\mathcal{N}=2$ super Yang-Mills theory in 
the $\Omega$-background and discuss the deformed supersymmetry.
In section 3, 
we calculate the central charges of the
supersymmetry algebra and derive the monopole equation
in the $\Omega$-background
by the Bogomol'nyi completion of the energy density.
We also discuss the 
supersymmetries preserved by the BPS equation.
Section 4 is devoted to conclusions and discussion.

%%%%%%%%%%%%%Section 2%%%%%%%%%%%%%%

\section{Supersymmetries of $\Omega$-deformed $\mathcal{N}=2$ super Yang-Mills theory}
In this section we discuss the deformation of 
$\mathcal{N}=2$ $U(N)$ super Yang-Mills theory 
in the $\Omega$-background \cite{Nekrasov:2002qd, Losev:2003py, NeOk, ItNaSaSa}. 
We will define the theory in spacetime with Euclidean signature.
The theory contains a gauge field $A_{m}$ ($m=1,2,3,4$), 
Weyl fermions $\Lambda^{I}_{\alpha}$, $\bar{\Lambda}^{I}_{\dot{\alpha}}$, 
and complex scalars $\varphi$, $\bar{\varphi}$. They belong to 
the adjoint representation of $U(N)$ gauge group. 
The $SO(4)=SU(2)_{L}\times SU(2)_{R}$ Lorentz spinor indices 
are denoted as $\alpha,\ \dot{\alpha}=1,2$ while 
$I=1,2$ indicates the $SU(2)_{I}$ R-symmetry index.
These $SU(2)$ indices are raised and lowered 
by the antisymmetric $\epsilon$-symbol 
with $\epsilon^{12}=-\epsilon_{12}=1$. 
We expand the fields with $U(N)$
basis $T^{u}$ $(u=1,2,\ldots,N^{2})$ normalized by
$\mathrm{Tr}(T^{u}T^{v})=\kappa\delta^{uv}$
with a certain constant $\kappa$. 
The Lagrangian of the theory in the flat spacetime
is given by 
\begin{align}
\mathcal{L}_0
&=
\frac{1}{\kappa}\textrm{Tr}\biggl[
\frac{1}{4}F_{mn}F^{mn}
-\frac{i\theta g^{2}}{32\pi^{2}}F_{mn}\tilde{F}^{mn}
+\Lambda^{I}\sigma^{m}D_{m}\bar{\Lambda}_{I}
+D_{m}\varphi D^{m}\bar{\varphi}
\notag\\
&\qquad\qquad{}
-i\frac{g}{\sqrt{2}}\Lambda^{I}[\bar{\varphi},\Lambda_{I}]
+i\frac{g}{\sqrt{2}}\bar{\Lambda}_{I}[\varphi,\bar{\Lambda}^{I}]
+\frac{g^{2}}{2}[\varphi,\bar{\varphi}]^{2}\biggr],
\label{undef}
\end{align}
where $F_{mn}=\partial_{m}A_{n}-\partial_{n}A_{m}+ig[A_{m},A_{n}]$ 
is the gauge field strength, 
$g$ is the gauge coupling constant and $D_{m}=\partial_{m}+ig[A_{m},\ast]$ 
is the gauge covariant derivative. 
We also define the Dirac matrices 
$\sigma_{m}=(i\tau^1,i\tau^2,i\tau^3,1)$ and
$\bar{\sigma}_{m}=(-i\tau^1,-i\tau^2,-i\tau^3,1)$,
where $\tau^{c}$ ($c=1,2,3$) are the Pauli matrices. 
The constant 
$\theta$ is the theta-angle and 
$\tilde{F}_{mn}=\frac{1}{2}\epsilon_{mnpq}F^{pq}$ is the dual of $F_{mn}$.

The four-dimensional $\mathcal{N} = 2$ super Yang-Mills theory 
in the $\Omega$ background is obtained by the dimensional reduction of
the six-dimensional $\mathcal{N} = 1$ super Yang-Mills theory with
the non-trivial metric \cite{NeOk}.
We also introduce the R-symmetry Wilson line by gauging the 
$SU(2)_{I}$ R-symmetry.
The Lagrangian of the four-dimensional theory is given by 
\begin{align}
\mathcal{L}^{}_{\Omega}
&=
\frac{1}{\kappa}\textrm{Tr}\biggl[
\frac{1}{4}F_{mn}F^{mn}
-\frac{i\theta g^{2}}{32\pi^{2}}F_{mn}\tilde{F}^{mn}
+(D_{m}\varphi-gF_{mn}\Omega^{n})
(D^{m}\bar{\varphi}-gF^{mp}\bar{\Omega}_{p})
\notag\\
&\qquad\qquad{}
+\Lambda^{I}\sigma^{m}D_{m}\bar{\Lambda}_{I}
-\frac{i}{\sqrt{2}}g\Lambda^{I}[\bar{\varphi},\Lambda_{I}]
+\frac{i}{\sqrt{2}}g\bar{\Lambda}_{I}[\varphi,\bar{\Lambda}^{I}]
\notag\\
&\qquad\qquad{}
+\frac{1}{\sqrt{2}}g\bar{\Omega}^{m}\Lambda^{I}
D_{m}\Lambda_{I}
-\frac{1}{2\sqrt{2}}g\bar{\Omega}^{mn}\Lambda^{I}
\sigma_{mn}\Lambda_{I}
\notag\\
&\qquad\qquad{}
-\frac{1}{\sqrt{2}}g\Omega^{m}\bar{\Lambda}_{I}
D_{m}\bar{\Lambda}^{I}
+\frac{1}{2\sqrt{2}}g\Omega^{mn}\bar{\Lambda}_{I}
\bar{\sigma}_{mn}\bar{\Lambda}^{I}
\notag\\
&\qquad\qquad{}
+\frac{g^{2}}{2}\Bigl([\varphi,\bar{\varphi}]
+i\Omega^{m}D_{m}\bar{\varphi}-i\bar{\Omega}^{m}D_{m}\varphi
+ig\bar{\Omega}^{m}\Omega^{n}F_{mn}
\Bigr)^{2}
\notag\\
&\qquad\qquad{}
-\frac{1}{\sqrt{2}}g\bar{\mathcal{A}}^{J}{}_{I}
\Lambda^{I}\Lambda_{J}
-\frac{1}{\sqrt{2}}g\mathcal{A}^{J}{}_{I}
\bar{\Lambda}^{I}\bar{\Lambda}_{J}
\biggr],
\label{omega_lag}
\end{align}
where the Lorentz generators $\sigma^{mn}$ and $\bar{\sigma}^{mn}$ are defined
by
\begin{gather}
(\sigma^{mn})_{\alpha}{}^{\beta}=
\frac{1}{4}\bigl(
\sigma^{m}_{\alpha\dot{\alpha}}\bar{\sigma}^{n\dot{\alpha}\beta}
-\sigma^{n}_{\alpha\dot{\alpha}}\bar{\sigma}^{m\dot{\alpha}\beta}
\bigr),
\quad 
(\bar{\sigma}^{mn})^{\dot{\alpha}}{}_{\dot{\beta}}=
\frac{1}{4}\bigl(
\bar{\sigma}^{m\dot{\alpha}\alpha}\sigma^{n}_{\alpha\dot{\beta}}
-\bar{\sigma}^{n\dot{\alpha}\alpha}\sigma^{m}_{\alpha\dot{\beta}}
\bigr).
\end{gather}
The R-symmetry Wilson line gauge fields $\mathcal{A}^{I}{}_{J}$,
$\bar{\mathcal{A}}^{I}{}_{J}$ are constant, 
$\Omega^m = \Omega^{mn} x_n, \bar{\Omega}^{m} = \bar{\Omega}^{mn}
x_n$ and the $\Omega$-background is
parametrized as follows:
\begin{gather}
\Omega^{mn}=\frac{1}{2\sqrt{2}}
\begin{pmatrix}
0 & i\epsilon_{1} & 0 & 0  \\
-i\epsilon_{1} & 0 & 0 & 0 \\
0 & 0 & 0 & -i\epsilon_{2} \\
0 & 0 & i\epsilon_{2} & 0  \\
\end{pmatrix}, 
\quad 
\bar{\Omega}^{mn}=\frac{1}{2\sqrt{2}}
\begin{pmatrix}
0 & -i\bar{\epsilon}_{1} & 0 & 0  \\
i\bar{\epsilon}_{1} & 0 & 0 & 0 \\
0 & 0 & 0 & i\bar{\epsilon}_{2} \\
0 & 0 & -i\bar{\epsilon}_{2} & 0  \\
\end{pmatrix}.
\label{omega}
\end{gather}
Here
$\epsilon_1$ and $\epsilon_2$ are complex numbers
and $\bar{\epsilon}_1$ and $\bar{\epsilon}_2$ are 
their complex conjugates.

The undeformed theory defined by \eqref{undef}
has $\mathcal{N}=2$ supersymmetry. The 
supersymmetry algebra is given by
\begin{eqnarray}
\begin{aligned}
\{Q_\alpha{}^I, \bar{Q}_{\dot{\beta}J}\}
&=
2 \left( \sigma^m \right)_{\alpha \dot{\beta}} P_m \delta^I{}_J,
\\
\{Q_{\alpha I}, Q_{\beta J}\}
&=
- 2 \sqrt{2} \epsilon_{IJ} \epsilon_{\alpha \beta}Z, 
\\
 \{\bar{Q}_{\dot{\alpha}}{}^I, \bar{Q}_{\dot{\beta}}{}^J\}
&=
- 2 \sqrt{2} \epsilon^{IJ} \epsilon_{\dot{\alpha} \dot{\beta}}\bar{Z},
\label{SUSYalgebra}
\end{aligned}
\end{eqnarray}
where $Q^I_\alpha$ and $\bar{Q}_I^{\dot{\alpha}}$ are
supercharges, $P^m$ is the four-momentum and $Z$ is the central charge.
The supersymmetry algebra leads to the BPS inequality for mass $M$:
\begin{eqnarray}
M \geq \sqrt{2} |Z|.
\label{BPSmass}
\end{eqnarray}
Here the equality holds for the BPS saturated states.

It is convenient to use the topological
twist \cite{Wi}, which is defined by 
identifying the $SU(2)_I$ R-symmetry indices $I$
with the $SU(2)_R$ spinor indices $\dot{\alpha}$. 
The twisted supercharges
are introduced by
\begin{gather}
Q_{m}=\bar{\sigma}_{m}^{I\alpha}Q_{\alpha I}, 
\quad 
\bar{Q}=\delta^{\dot{\alpha}}_{I}\bar{Q}_{\dot{\alpha}}^{I}, 
\quad 
\bar{Q}_{mn}=-(\bar{\sigma}_{mn})^{\dot{\alpha}}{}_{I}
\bar{Q}_{\dot{\alpha}}^{I}.
\label{twisted_supercharges}
\end{gather}
The supersymmetry algebra \eqref{SUSYalgebra} becomes
\begin{align}
\begin{aligned}
\{Q_m, \bar{Q}\} = 2P_m,
&\quad
\label{twisusy2}
\{Q_m, \bar{Q}_{pq} \}
=
\left(
\delta^{nq}\delta^{mp} - \delta^{np} \delta^{mq} - \varepsilon^{mnpq}
\right) P_n,
\\
\{ \bar{Q} , \bar{Q} \} =
4 \sqrt{2}\bar{Z},
&\quad
\{Q_m, Q_n \} 
=
- 4 \sqrt{2}\delta_{mn}Z,
\\
\{ \bar{Q}_{mn}, \bar{Q} \}=0,
&\quad
\{\bar{Q}_{mn}, \bar{Q}_{pq} \}
=
\frac{\sqrt{2}}{4}\left(
\delta^{mp}\delta^{nq} - \delta^{mq}\delta^{np} - \varepsilon^{mnpq}
\right)
\bar{Z}.
\end{aligned}
\end{align}

For general $\Omega$-background, where $\epsilon_1$ and
$\epsilon_2$ are generic, the action defined by $\eqref{omega_lag}$
has one scalar supersymmetry $\bar{Q}$ \cite{Nekrasov:2002qd, ItNaSaSa} by choosing the R-symmetry
Wilson line gauge fields such as
\begin{gather}
\mathcal{A}^{I}{}_{J}
=-\frac{1}{2}\Omega_{mn}(\bar{\sigma}^{mn})^{I}{}_{J},\quad 
\bar{\mathcal{A}}^{I}{}_{J}
=-\frac{1}{2}\bar{\Omega}_{mn}(\bar{\sigma}^{mn})^{I}{}_{J}. 
\label{condition1}
\end{gather}
If the $\Omega$-background and the Wilson line
satisfy special conditions, the theory has further extended
supersymmetries.

It is known that if the $\Omega$-background are
self-dual
$\epsilon_1 + \epsilon_2 =0$ or anti-self-dual
$\epsilon_1 - \epsilon_2 = 0$
and there are no Wilson lines, the theory 
has $\mathcal{N}=(4,0)$ or $\mathcal{N}=(0,4)$
supersymmetries\footnote{We denote
$\mathcal{N}=(p,q)$  by supersymmetry with $p$ chiral and $q$
anti-chiral supercharges.} \cite{Ito:2008qe}. The
deformed supersymmetry transformations are given by
\begin{align}
\begin{aligned}
\delta A_m =&
- \left( \xi^I \sigma_m \bar{\Lambda}_I + \bar{\xi}_I \bar{\sigma}_m \Lambda^I \right), \\
\delta \Lambda^I = &
\sigma^{mn} \xi^I F_{mn}
+ \sqrt{2} \sigma^m \bar{\xi}^I D_m \varphi 
+ ig \xi^I \left[ \varphi, \bar{\varphi} \right] \\
&-\sqrt{2} g F_{mn} \Omega^n \sigma^m \bar{\xi}^I
+ g \xi^I \left( \bar{\Omega}^m D_m \varphi - \Omega^m D_m \bar{\varphi} \right)
+ g^2 \Omega^m \bar{\Omega}^n F_{mn} \xi^I, \\
\delta \bar{\Lambda}_I =&
\bar{\sigma}^{mn} \bar{\xi}_I F_{mn} 
- \sqrt{2} \bar{\sigma}^m \xi_I D_m \bar{\varphi}
- ig \bar{\xi}_I \left[ \varphi, \bar{\varphi} \right]  \\
&- \sqrt{2} g F_{mn} \bar{\Omega}^n \xi_I \sigma^m  
- g \bar{\xi}_I \left( \bar{\Omega}^m D_m \varphi 
- \Omega^m D_m \bar{\varphi} \right) 
+ g^2 \bar{\Omega}^m \Omega^n F_{mn} \bar{\xi}_I, \\
\delta \varphi =&
\sqrt{2} \xi^I \Lambda_I 
- g \Omega^m \left( \xi^I \sigma_m \bar{\Lambda}_I  
+ \bar{\xi}_I \bar{\sigma}_m\Lambda^I \right), \\
\delta \bar{\varphi} =&
\sqrt{2} \bar{\xi}^I \bar{\Lambda}_I 
- g \bar{\Omega}^m \left( \xi^I \sigma_m \bar{\Lambda}_I  
+ \bar{\xi}_I \bar{\sigma}_m\Lambda^I \right).
\end{aligned}\label{defsusy}
\end{align}
Here, we have set $\xi^I = 0$ for the self-dual
and $\bar{\xi}^I = 0$ for
the anti-self-dual $\Omega$-background.
We note that in these cases, there are no
translational symmetries and
only the chiral- or anti-chiral-
sector of the supersymmetry remains.

We now discuss supersymmetries of the theory
with the Wilson line (\ref{condition1})
in the cases that one of the deformation parameters $\epsilon_1,
\epsilon_2$ is zero.
In these cases, the translational invariance in the $(1,2)$-plane or the
$(3,4)$-plane is restored.

As we have mentioned before,
at least one supersymmetry $\bar{Q}$ corresponding to
the transformation with $\bar{\xi} = \delta^{\dot{\alpha}}_I \bar{\xi}^I_{\dot{\alpha}}$ 
in \eqref{defsusy}
is conserved.
We examine invariance of the action under other transformations
generated by supercharges $\bar{Q}^{mn}$ and $Q^m$
associated with the deformed supersymmetry transformations
\eqref{defsusy}.
We find that the action is not invariant by the
transformation generated by $\bar{Q}^{mn}$.
The variation of the Lagrangian under the transformation generated by $Q^m$ is 
\begin{align}
\delta_{\xi_m} \mathcal{L} & =
\xi_m \left(\bar{\Omega}^{+mn} + \bar{\Omega}^{-mn}\right) 
\left\{ \sqrt{2} g \delta_{np} F^{+pq} \Lambda_q 
- \frac{1}{\sqrt{2}} i g^2 \left[ \varphi , \bar{\varphi} \right]  \Lambda_n
\right. \notag \\
&
\left.
+ \frac{1}{\sqrt{2}} g^2 \Lambda_n
\left( \Omega^p D_p \bar{\varphi} - \bar{\Omega}^p D_p \varphi \right)
 -\frac{1}{\sqrt{2}} g^3 \Omega^p \bar{\Omega}^q F_{pq} \Lambda_n
\right\} 
\notag \\
& +\xi_m \left(\Omega^{+mn} + \Omega^{-mn}\right)\left\{
 g \left(D^p \varphi \right)
\left( 2 \bar{\Lambda}^-_{np} + \delta_{np} \bar{\Lambda} \right) 
+g^2\bar{\Omega}^q \left( 2 F_{pq}\bar{\Lambda}^-_{pn} + F_{nq} \bar{\Lambda}\right)  
\right\},
\label{var1}
\end{align}
where $\Lambda_m, \bar{\Lambda}_{mn}, \bar{\Lambda}$ 
are the
fermions obtained by the twist \eqref{twisted_supercharges}.
The (anti-)self-dual part of an
antisymmetric tensor $A_{mn}$ 
is defined by $A_{mn}^{\pm}=\frac{1}{2}(A_{mn}\pm\tilde{A}_{mn})$.

{}From \eqref{var1}, we find that the variation by the $Q^m$-transformation vanishes
under the conditions:
\begin{align}
\xi_m \left( \Omega^{+mn} + \Omega^{-mn} \right)= 0 ,
\quad
\xi_m \left( \bar{\Omega}^{+mn} + \bar{\Omega}^{-mn} \right) = 0.
\end{align}
When $\epsilon_2 = 0$, these conditions are satisfied for
$
\xi^m = \left(0, 0, \xi^3, \xi^4 \right).
$
Therefore for $\epsilon_2=0$, $\mathcal{N} = (0,1)$
supersymmetry is enhanced to $\mathcal{N} = (2,1)$
generated by supercharges $Q^3,\ Q^4,\ \bar{Q}$.
When $\epsilon_1 = 0$, these conditions are satisfied for
$
\xi^m = \left(\xi^1, \xi^2, 0, 0\right).
$
Therefore for $\epsilon_1=0$, 
the theory has $\mathcal{N} = (2,1)$ supersymmetry
generated by supercharges $Q^1,\ Q^2,\ \bar{Q}$.

We note that when we exchange
$\Omega^{mn}$ with $\bar{\Omega}^{mn}$
in the definition of the Wilson line \eqref{condition1},
we find that certain linear combination of
$\bar{Q}^{mn}$ is conserved. In this case,
the Wilson line becomes
\begin{gather}
\mathcal{A}^{I}{}_{J}
=-\frac{1}{2}\bar{\Omega}_{mn}(\bar{\sigma}^{mn})^{I}{}_{J},\quad 
\bar{\mathcal{A}}^{I}{}_{J}
=-\frac{1}{2}\Omega_{mn}(\bar{\sigma}^{mn})^{I}{}_{J}.
\label{condition2}
\end{gather}
We find that the action is no longer invariant by the
transformation generated by $\bar{Q}$.
The variations of the Lagrangian by $Q^m$ and $\bar{Q}^{mn}$ become
\begin{align}
\delta_{\xi_m} \mathcal{L} & =
\xi_m \left(\bar{\Omega}^{+mn} + \Omega^{-mn}\right) 
\left\{ \sqrt{2} g \delta_{np} F^{+pq} \Lambda_q 
- \frac{1}{\sqrt{2}} i g^2 \left[ \varphi , \bar{\varphi} \right]  \Lambda_n
\right. \notag \\
&
\left.
+ \frac{1}{\sqrt{2}} g^2 \Lambda_n \left( \Omega^p D_p \bar{\varphi}
- \bar{\Omega}^p D_p \varphi \right)
-\frac{1}{\sqrt{2}} g^3 \Omega^p \bar{\Omega}^q F_{pq}  \Lambda_n
\right\} 
\notag \\
& +\xi_m \left(\Omega^{+mn} + \bar{\Omega}^{-mn}\right)\left\{
 g \left(D^p \varphi \right)
\left( 2 \bar{\Lambda}^-_{np} + \delta_{np} \bar{\Lambda} \right) 
+g^2\bar{\Omega}^q \left( 2 F_{pq}\bar{\Lambda}^-_{pn} + F_{nq} \bar{\Lambda}\right)  
\right\},
\label{var2}
\end{align}
\begin{align}
\delta_{\bar{\xi}_{mn}}\mathcal{L} &=  
 \bar{\xi}_{mn} \left(  \Omega^{-pm}+ \bar{\Omega}^{-pm}\right)
 \frac{1}{\sqrt{2}} g \Big\{
 2 F^-_{qp}  \bar{\Lambda}^-_{nq} - F^-_{np} \bar{\Lambda} 
 + ig\bar{\Lambda}_{pn} \left[ \varphi, \bar{\varphi} \right]
 \notag \\
 &
 + g \left( \Omega^q D_q \bar{\varphi} - \bar{\Omega}^q D_q \varphi \right)\bar{\Lambda}^-_{np}
 + g^2 \bar{\Omega}^{l} \Omega^q F_{lq} \bar{\Lambda}^-_{np}
 -2\sqrt{2}g \Omega^q F_{pq} \Lambda_n 
 +2\sqrt{2} D_p \varphi \Lambda_n 
  \Big\}
 \notag \\
 &+ \bar{\xi}_{mn}\Omega^{-mn} \left(
  \frac{1}{\sqrt{2}} g  F^-_{pq} \bar{\Lambda}^{-pq} 
 + g\left( D_p \varphi \right) \Lambda_p \right)
 + \bar{\xi}_{mn} g^2 \Omega^{-mn} F_{pq} \Omega^p \Lambda^q 
 \notag \\
 &
 +\bar{\xi}_{mn} \left( \Omega^{-mn} -
 \bar{\Omega}^{-mn} \right)\frac{1}{2\sqrt{2}}g^2
 \left\{
 -i \left[ \varphi, \bar{\varphi} \right] \bar{\Lambda} 
 + \left( \Omega^p D_p \bar{\varphi} - \bar{\Omega}^p D_p \varphi \right) \bar{\Lambda}
 + g \bar{\Omega}^{p} \Omega^q F_{pq} \bar{\Lambda}\right\}.
 \label{var3}
\end{align}
These vanish for
\begin{align}
\begin{aligned}
\bar{\xi}^{mn}\left( \Omega^-_{pm}+ \bar{\Omega}^-_{pm}  \right) = 0,
& \quad 
\bar{\xi}^{mn} \Omega^-_{mn}  = 0, \\
\xi^m \left( \bar{\Omega}^-_{mn} + \Omega^+_{mn} \right)  = 0, 
& \quad
\xi^m \left( \Omega^-_{mn} + \bar{\Omega}^+_{mn} \right)  = 0.
\end{aligned}
\label{susy_preserving_condition}
\end{align}
When $\epsilon_2=0$, these conditions are satisfied for
real $\epsilon_1$, $\xi^m = ( \xi^1, \xi^2, 0, 0)$ and 
\begin{align}
 \bar{\xi}^{mn} & =
\left(
\begin{array}{cccc}
0  & 0 &   \bar{\xi}_{13} & \bar{\xi}_{14} \\ 
0  & 0 &  -\bar{\xi}_{14} & \bar{\xi}_{13} \\
-\bar{\xi}_{13} & \bar{\xi}_{14} &  0 & 0 \\
-\bar{\xi}_{14} & -\bar{\xi}_{13} &  0 & 0
\end{array} \right).
\label{barximn}
\end{align}
Therefore for $\epsilon_2=0$, the theory has
$\mathcal{N} = (2,2)$ supersymmetry generated by
supercharges $Q^1,\ Q^2,\ \bar{Q}^{13},\ \bar{Q}^{14}$.
When $\epsilon_1=0$, these conditions are satisfied for
real $\epsilon_2$,
$
\xi^m = ( 0, 0, \xi^3, \xi^4),
$
and $\bar{\xi}^{mn}$ given by \eqref{barximn}.
Therefore for $\epsilon_1=0$, the theory has 
$\mathcal{N} = (2,2)$ supersymmetry and the supercharges
$Q^3,\ Q^4,\ \bar{Q}^{13},\ \bar{Q}^{14}$
are conserved.

The conserved supercharges are summarized in table
\ref{preserved_SUSY}.
\begin{table}[t]
\begin{center}
\begin{tabular}{|l||c|c|}
\hline
& $\epsilon_1 \not=0, \ \epsilon_2 = 0$ & $\epsilon_1 = 0, \ \epsilon_2 \not=0$ \\
\hline
\hline
Wilson line \eqref{condition1} 
& 
$Q^3, \ Q^4, \ \bar{Q}$
& 
$Q^1, \ Q^2, \ \bar{Q}$
\\
\hline
Wilson line \eqref{condition2} 
& 
$Q^1, \ Q^2, \ \bar{Q}^{13}, \ \bar{Q}^{14}$
& 
$Q^3, \ Q^4, \ \bar{Q}^{13}, \ \bar{Q}^{14}$
\\
\hline
\end{tabular}
\caption{Classification of conserved supercharges.}
\label{preserved_SUSY}
\end{center}
\end{table}
In all cases, the theory has supersymmetries including
both the chiral- and anti-chiral-sectors.
So far we have identified the $SU(2)_I$ R-symmetry
indices with the $SU(2)_R$ spinor indices.
When we identify them with the $SU(2)_L$
spinor indices $\alpha$ and replace $\bar{\sigma}^{mn}$
in the Wilson lines by $\sigma^{mn}$,
the theory has $\mathcal{N}=(1,2)$ or $\mathcal{N}=(2,2)$
supersymmetry.

%%%%%%%%%%%%%Section 3%%%%%%%%%%%%%%
\section{Central charges and BPS monopole equation}
In this section, we will derive the Noether currents and 
evaluate the central charge of the algebra.
We will obtain the BPS monopole equation and classify 
supersymmetries preserved by the equation, which depend on the 
$\Omega$-background and the $SU(2)_I$ Wilson line.
{}From the deformed supersymmetry transformations 
\eqref{defsusy}, we can calculate
the Noether currents $J^{mI}_\alpha$ and $\bar{J}_I^{m\dot{\alpha}}$
associated with them. 
We find
\begin{align}
J^m_{I \alpha} = & \frac{1}{\kappa} \mathrm{Tr} \Bigg[
\sqrt{2} \left( D_m \bar{\varphi} \right) \Lambda_{I \alpha}
+ \left\{  
-ig \left[ \varphi, \bar{\varphi} \right] \delta_{mn} 
- F_{mn} 
+ \left( \frac{i \theta g^2 }{8 \pi^2 } + 1 \right)\tilde{F}_{mn} 
\right\}
\sigma^n_{\alpha \dot{\alpha}} \bar{\Lambda}_I{}^{\dot{\alpha}} \notag \\
&-2\sqrt{2} \left( D_n \bar{\varphi} \right)  \sigma^{mn}{}_\alpha {}^\beta \Lambda_{I \beta }
- g \left( \varphi \bar{\Omega}^m - \bar{\varphi} \Omega^m \right)
\sigma^n_{\alpha \dot{\alpha} } D_n \bar{\Lambda}_I{}^{\dot{\alpha}}
-g \left( \varphi \bar{\Omega}_{mn} - \bar{\varphi} \Omega_{mn} \right)
\sigma^n_{\alpha \dot{\alpha}} \bar{\Lambda}_I{}^{\dot{\alpha}} \notag\\
&+g \left( \varphi \bar{\Omega}_n - \bar{\varphi} \Omega_n \right)
\sigma^m_{\alpha \dot{\alpha}} D^n \bar{\Lambda}_I{}^{\dot{\alpha}} 
- g\left(  \Omega^m D^n\bar{\varphi} +  \bar{\Omega}^m D^n \varphi \right)
\sigma^n_{\alpha \dot{\alpha}} \bar{\Lambda}_I{}^{\dot{\alpha}} \notag \\
&-\sqrt{2} g F_{mn} \bar{\Omega}^n  \Lambda_{I\alpha}
- i \sqrt{2} g^2 \bar{\Omega}^m \left[ \varphi, \bar{\varphi} \right]   \Lambda_{I\alpha} 
+   \sqrt{2} g^2 \bar{\Omega}^m \left( \Omega^n D_n \bar{\varphi}
- \bar{\Omega}^n D_n \varphi \right)  \Lambda_{I\alpha}  \nonumber \\
&  + 2 \sqrt{2} g F_{np} \bar{\Omega}^p \sigma^{mn}{}_\alpha {}^\beta \Lambda_{I \beta}
+ \sqrt{2} gF_{np} \bar{\Omega}^m  \sigma^{np} {}_\alpha {}^\beta \Lambda_{I\beta} \nonumber \\
& + 2g^2 F_{np} \Omega^m \bar{\Omega}^p \sigma^n_{\alpha \dot{\alpha}} \bar{\Lambda}_{I} {}^{\dot{\alpha}}
- g^2 \bar{\Omega}^p \Omega^n  F_{np} \sigma^m_{\alpha \dot{\alpha}} \bar{\Lambda}_I {}^{\dot{\alpha}} 
+ \sqrt{2} g^3 \bar{\Omega}^m \bar{\Omega}^n \Omega^p F_{np}  \Lambda_{I\alpha} \Bigg].
\label{supercurrent}
\end{align}
The complex conjugate $\bar{J}^{\dot{\alpha} m}_I$ may be
calculated in a similar way. 
The anti-commutation relations of 
the supercharges, which are defined by the spatial integration of 
$J^{4I}_{\alpha}$
is evaluated by using canonical anti-commutation relations of the
fermions, which is given by 
\begin{align}
\left\{ \bar{\Lambda}_I {}^{\dot{\alpha}}(\vec{x}, x^4) , \Lambda^{J\alpha}(\vec{x}', x^4) \right\} 
= \delta_I {}^J \bar{\sigma}^{4\dot{\alpha} \alpha} \delta^3 (\vec{x} - \vec{x}').
\end{align}
We then obtain \cite{Witten:1978mh}
\begin{align}
\left\{ Q_{I\alpha} , Q_{J\beta} \right\}
=  - 2\sqrt{2} \epsilon_{\alpha \beta} \epsilon_{IJ} Z,
\end{align}
where $Z$ is the deformed central charge:
\begin{align}
Z &=  
\int d^3 x \frac{1}{\kappa} \mathrm{Tr}
\Bigg[  ig \left( D_4 \bar{\varphi} \right)\left[ \varphi, \bar{\varphi} \right]
+ \left( D^n \bar{\varphi} \right) 
\left\{   F_{4n} - \left( \frac{i \theta g^2 }{8 \pi^2 } + 1 \right)\tilde{F}_{4n}\right\}  \nonumber \\
& + g \left( D_4 \bar{\varphi} \right) \left( \bar{\Omega}^n D_n \varphi - \Omega^n D_n \bar{\varphi} \right)
+ g \left\{  -ig \left[ \varphi, \bar{\varphi} \right] \delta_{4n} 
- F_{4n} + \left( \frac{i \theta g^2 }{8 \pi^2 } + 
1 \right)\tilde{F}_{4n}  \right\} F_{np} \bar{\Omega}^p \nonumber \\
& + g^2 \left( D_4 \bar{\varphi} \right) \bar{\Omega}^p \Omega^n F_{np}
- g^2 \left( \bar{\Omega}^n D_n \varphi - 
\Omega^n D_n \bar{\varphi} \right) F_{4p}\bar{\Omega}^p 
- g^3 F_{4n} \bar{\Omega}^n \bar{\Omega}^p \Omega^q F_{qp } \Bigg].
\end{align}
The complex conjugate $\bar{Z}$ can be calculated in a similar way.
In this paper, we focus on the magnetic monopole configuration
such that
the fields that depend on the three-dimensional space spanned by 
$(x^1,x^2,x^3)$ and are independent of $x^4$. 
The scalar field is taken to be 
in the Cartan subalgebra so that $[\varphi, \bar{\varphi}]=0$
and we fix the gauge $A_4 = 0$ so that the
electric field $F_{4i}$ vanishes.
We find that $Z$ for the monopole configuration is given by,
\begin{equation}
Z = - \left(  \frac{i \theta g^2 }{8 \pi^2 } + 1 \right) 
\frac{1}{\kappa} \int \!  d^3 x \ \partial^i 
\mathrm{Tr} [ B_i \varphi ],
\label{cc}
\end{equation}
where we have defined the magnetic field 
$B_i = \frac{1}{2} \epsilon_{ijk} F^{jk}$.
This is the same as the undeformed case. 
However, we will find that the monopole equation is deformed by the
$\Omega$-background as discussed below. Then the central charge could
depend on the $\epsilon$-parameter.

To find the BPS equation corresponding to the monopole configuration, 
we perform the Bogomol'nyi completion of the energy functional by 
combining the kinetic and potential terms.
In the monopole configuration, we obtain the energy 
\begin{eqnarray}
E &=& \frac{1}{\kappa} \int \! d^3 x \ \mathrm{Tr}
\left[
\frac{1}{2} B_i^2 + (D_i \varphi + g \Omega^j F_{ji}) (D_i \bar{\varphi}
+ g \bar{\Omega}_k F^{ki}) 
\right. 
\nonumber \\
& & \qquad 
\left.
+ \frac{g^2}{2} 
\left(
[\varphi, \bar{\varphi}] + i \Omega^i D_i \bar{\varphi}
- i \bar{\Omega}^i D_i \varphi - i g \Omega^i \bar{\Omega}^j F_{ij}
\right)^2
\right],
\label{energy}
\end{eqnarray}
where we have taken $\theta = 0$ for simplicity.
The vacuum configurations of the theory are given by
\begin{align}
D_i \varphi = D_i \bar{\varphi} = 0,
\quad
A_i = 0,
\quad
\left[ \varphi, \bar{\varphi} \right] = 0.
\end{align}
Hence, the Higgs field $\varphi$ takes value 
in the Cartan subalgebra $U(1)^N$ 
and the vacuum moduli space becomes
$U(N)/U(1)^N \cong SU(N)/U(1)^{N-1}$, which is
the same as the undeformed theory.

To find the energy bound, we use the phase transformation of $\varphi$
and $\Omega$ to set those to 
the values which are consistent with 
the first equation in \eqref{susy_preserving_condition}:
\begin{eqnarray}
\varphi = - \bar{\varphi}, \quad \Omega^{mn} = - \bar{\Omega}^{mn}.
\label{BPS_field}
\end{eqnarray}
Then the third term in \eqref{energy} vanishes and 
the energy is rewritten in the perfect square
form,
\begin{eqnarray}
E &=& \frac{1}{\kappa} \int \! d^3 x \ \mathrm{Tr}
\left[
\frac{1}{2} 
\left(
B_i \pm (D_i \phi + g \hat{\Omega}^j F_{ji})
\right)^2 
\right] 
\mp \frac{1}{\kappa} 
\int \! d^3 x \ \partial_i 
\mathrm{Tr} 
\left[
B_i \phi
\right] 
\nonumber \\
&\ge& 
\mp \frac{1}{\kappa} 
\int \! d^3 x \ \partial_i 
\mathrm{Tr} 
\left[
B_i \phi
\right],
\label{Bcompenergy}
\end{eqnarray}
where we have defined $\phi = i\sqrt{2} \varphi, \ 
\hat{\Omega}^{ij} = i\sqrt{2} \Omega^{ij}$ and $\hat{\Omega}^i
= \hat{\Omega}^{in} x_n$. 
The energy bound is saturated if the following deformed BPS equation is satisfied:
\begin{align}
B_i \pm (D_i \phi + g \hat{\Omega}^j F_{ji}) = 0.
\label{BPSeq}
\end{align}
{}From \eqref{Bcompenergy}, we find that the 
energy at the lower bound is the same as 
the central charges \eqref{cc} derived from the supersymmetry
algebra. 

When $\epsilon_1 \not=0$ and $\epsilon_2 = 0$, 
the BPS equation is deformed by $\epsilon_1$ and 
its solutions depend on the parameter $\epsilon_1$.
On the other hand, when $\epsilon_1 =0$ and $\epsilon_2 \not= 0$,
the BPS equation depends on $x^4$
since the third term in the equation 
(\ref{BPSeq}) contains $x^4$.
In this case, we can consider the
monopole configuration which depends
on $(x^2,x^3,x^4)$ and is independent of
$x^1$, which gives the similar
deformed BPS equation.

In the following, we will investigate
supersymmetries that are preserved by the BPS state.
Substituting the BPS equation \eqref{BPSeq} into 
the supersymmetry transformations of the fermions in \eqref{defsusy}, we
obtain the condition on the supersymmetry parameters that are preserved
by the BPS configurations:
\begin{equation}
\pm i(\bar{\sigma}^{mn})^I {}_{\dot{\beta}} 
\varepsilon^{\dot{\beta} \alpha} \bar{\xi}_{mn} \pm 
i \varepsilon^{I \dot{\alpha}} \bar{\xi} +
\xi_m (\sigma^m)_{\alpha J} \varepsilon^{IJ} = 0, \quad
(\alpha = \dot{\alpha} = 1,2).
\label{BPScondition}
\end{equation}
As we have discussed in section 2, 
only part of the $\xi_m, \bar{\xi}, \bar{\xi}_{mn}$ 
symmetries exist in the theory
in the $\Omega$-background.
We will examine the condition \eqref{BPScondition} for the cases where
$\Omega^{mn}$ is (anti-)self-dual or one of the
$\epsilon$-parameters is zero.

When $\Omega^{mn}$ and $\bar{\Omega}^{mn}$ are 
(anti-)self-dual, the action is invariant under 
$\xi_{\alpha I}$ (anti-self-dual case) or
$\bar{\xi}_{\dot{\alpha} I}$ (self-dual case) transformations.
In both cases, we find that all the supersymmetry is broken by the 
BPS conditions \eqref{BPScondition}.

On the other hand, when $\Omega^{mn}$ and $\bar{\Omega}^{mn}$ are not (anti-)self-dual, 
the action is invariant under the $\bar{\xi}$ 
transformation if one considers the Wilson line \eqref{condition1}.
The condition \eqref{BPScondition} implies that the $\bar{\xi}$-supersymmetry
is broken and the BPS configuration does not preserve any supersymmetries.

In the cases that one of
the $\epsilon$-parameters vanishes, the supersymmetries that are preserved by the BPS
condition are classified as follows:
%%%%%%%%%%%%%%%%%%%%%%%%%%%%%%%%%%%%%%%%%%%%%
\begin{enumerate}
\item \underline{$\epsilon_1 \not= 0, \ \epsilon_2 = 0$ and Wilson line \eqref{condition1}} \\
In this case, the action is invariant under the $\bar{\xi}, \xi_3,
\xi_4$ supersymmetries and there are translational 
symmetries in the (3,4)-plane. 
The BPS condition \eqref{BPScondition} becomes
\begin{align}
\left(
\begin{array}{cc}
0 & i \xi_3 - i \xi_4 \pm i\bar{\xi} \\
i \xi_3 + i \xi_4 \mp i\bar{\xi} & 0
\end{array}
\right) = 0,
\end{align}
where we have written the condition as the 
$2\times 2$ matrix form with respect 
to the spinor and R-symmetry indices.
Therefore the BPS state preserves one supersymmetry specified by 
a linear combination of $Q_4$ and $\bar{Q}$. 
%%%%%%%%%%%%%%%%%%%%%%%%%%%%%%%%%%%%%%%%%%%%%
\item \underline{$\epsilon_1 = 0, \ \epsilon_2 \not=0$ and 
Wilson line \eqref{condition1}} \\
In this case, the action is invariant under the 
$\bar{\xi}, \xi_1,\xi_2$ supersymmetries 
and the translational symmetry in the (1,2)-plane. 
The BPS condition \eqref{BPScondition} becomes
\begin{align}
\left(
\begin{array}{cc}
- i \xi_1 - \xi_2 & \pm i\bar{\xi} \\
\mp i \bar{\xi} & i \xi_1 - i \xi_2
\end{array}
\right) = 0.
\end{align}
This condition implies that $\xi_3 = \xi_4 = \bar{\xi} = 0$.
Therefore the BPS state does not preserve any supersymmetries.

%%%%%%%%%%%%%%%%%%%%%%%%%%%%%%%%%%%%%%%%%%%%%
\item \underline{$\epsilon_1 \not=0, \ \epsilon_2 = 0$ 
and Wilson line \eqref{condition2}} \\
In this case, the action is invariant under the 
$\xi_1, \xi_2, \bar{\xi}_{13}, \bar{\xi}_{14}$  supersymmetries
and the translational symmetry in the (3,4)-plane. 
The BPS condition \eqref{BPScondition} becomes
\begin{align}
\left(
\begin{array}{cc}
\pm i \bar{\xi}_{13} \mp \bar{\xi}_{14} - i \xi_1 - \xi_2 & 0 \\
0 & \pm i \bar{\xi}_{13} \pm  \bar{\xi}_{14} + i \xi_1 - \xi_2
\end{array}
\right) = 0.
\end{align}
This implies the conditions,
\begin{eqnarray}
\pm i\bar{\xi}_{13} - \xi_2 = 0, \quad 
\mp i\bar{\xi}_{14} + \xi_1 = 0.
\end{eqnarray}
Therefore the BPS state preserves two supersymmetries specified by linear
combinations of $Q_2, \bar{Q}_{13}$ and $Q_1, \bar{Q}_{14}$.

%%%%%%%%%%%%%%%%%%%%%%%%%%%%%%%%%%%%%%%%%%%%%
\item \underline{$\epsilon_1 = 0, \ \epsilon_2 \not= 0$ 
and Wilson line \eqref{condition2}} \\
In this case the action is invariant under the 
$\bar{\xi}_{13}, \bar{\xi}_{14}, \xi_3, \xi_4$  supersymmetries 
and the translational symmetry in the (1,2)-plane. 
The BPS condition \eqref{BPScondition} becomes
\begin{align}
\left(
\begin{array}{cc}
\pm i\bar{\xi}_{13} \mp  \bar{\xi}_{14} & i \xi_3 - i \xi_4 \\
i \xi_3 + i \xi_4 & \pm i\bar{\xi}_{13} \pm  \bar{\xi}_{14} 
\end{array}
\right)  = 0.
\end{align}
This condition implies $\bar{\xi}_{13} = \bar{\xi}_{14} = \xi_3 = \xi_4 = 0$.
Thus the BPS state does not preserve any supersymmetries.
\end{enumerate}
The supercharges that are preserved by the BPS condition 
are summarized in Table \ref{BPS_SUSY}.

\begin{table}[t]
\begin{center}
\begin{tabular}{|l||c|c|}
\hline
& $\epsilon_1 \not=0, \ \epsilon_2 = 0$ & $\epsilon_1 = 0, \ \epsilon_2 \not=0$ \\
\hline
\hline
Wilson line \eqref{condition1} 
& 
$iQ^4 \mp i\bar{Q}$
& 
no supercharge
\\
\hline
Wilson line \eqref{condition2} 
& 
$\pm i\bar{Q}^{13} - Q^2, \ \mp i\bar{Q}^{14} + Q^1$
& 
no supercharge
\\
\hline
\end{tabular}
\caption{Supercharges preserved by the BPS state.}
\label{BPS_SUSY}
\end{center}
\end{table}

%%%%%%%%%%%%%Section 4%%%%%%%%%%%%%%

\section{Conclusions and discussion}
In this paper, we have studied the deformed
supersymmetries 
of $\mathcal{N}=2$ super Yang-Mills theories 
in the $\Omega$-background and the Wilson lines.
For general $\Omega$-background, there is one scalar supersymmetry.
When one of the $\epsilon$-parameters is zero,
we have found the theories with $\mathcal{N}=(2,1)$ or $(2,2)$ or $(1,2)$
supersymmetry by choosing the appropriate Wilson line gauge fields. 
We have also calculated the central
charge of the deformed supersymmetry algebra.
For the monopole
configurations the formula for the central charge
does not contain the $\epsilon$-parameter.
We have performed the Bogomol'nyi completion of the energy density
and obtained the deformed BPS monopole equation.
We have examined the supersymmetries
preserved by the monopole equation.

The central charge for the monopole could receive
$\epsilon$-corrections through the deformed BPS monopole solution, where
this situation occurs for $\epsilon_2=0$.
In this case, the deformed BPS equation has the
axial symmetry around the $z$-axis.
We may use Manton's ansatz \cite{Ma} for the fields:
\begin{align}
\begin{aligned}
A^a_i&=\left\{ \eta_1\hat{\rho}^a+\left( \eta_2 + \frac{1}{g\rho} \right) \hat{z}^a \right\}\hat{\varphi}^i
+W_1\hat{\rho}^i\hat{\varphi}^a+W_2\hat{z}^i \hat{\varphi}^a,\\
\phi^a&=\phi_1\hat{\rho}^a+\phi_2\hat{z}^a,
\end{aligned}
\label{mansatz}
\end{align}
where $(\rho,\varphi,z)$ are the cylindrical coordinates.
$\eta_\alpha$, $W_\alpha$ and $\phi_\alpha$
($\alpha = 1,2$) are functions of
$(\rho, z)$ and 
\begin{align}
\hat{\rho}=(\cos\varphi,\sin\varphi,0),
\quad 
\hat{\varphi}=(-\sin\varphi,\cos\varphi,0),
\quad
\hat{z}=(0,0,1).
\end{align}
Here we have considered the
$SU(2)$ gauge group for simplicity and the
superscript $a$ denotes the $SU(2)$ index.
Substituting \eqref{mansatz} into the deformed BPS equation \eqref{BPSeq},
we obtain 
\begin{align}
\begin{aligned}
&
-\pa_3\eta_1+g\eta_2 W_2=
{1\over 1+g^2\epsilon^2 \rho^2}
\Bigl\{\pa_\rho\phi_1-gW_1\phi_2
+g\epsilon\rho (\pa_3\phi_1-g W_2\phi_2)
\Bigr\},
\\
&
-\pa_3\eta_2-g\eta_1 W_2=
{1\over 1+g^2\epsilon^2 \rho^2}
\Bigl\{
\pa_\rho\phi_2+gW_1\phi_1
+g\epsilon\rho(\pa_3\phi_2+g W_2\phi_1)
\Bigr\},
\\
&
\pa_\rho\eta_1+{\eta_1\over\rho}-g
W_1\eta_2=
{1\over 1+g^2\epsilon^2 \rho^2}
\Bigl\{
\pa_3\phi_1-g W_2\phi_2
-g\epsilon\rho (\pa_\rho\phi_1-g W_1\phi_2)
\Bigr\},
\\
&
\pa_\rho\eta_2+{\eta_2\over\rho}+g
W_1\eta_1=
{1\over 1+g^2\epsilon^2 \rho^2}
\Bigl\{
\pa_3\phi_2+g W_2\phi_1
-g\epsilon\rho (\pa_\rho\phi_2+g W_1\phi_1)
\Bigr\},
\\
&
\pa_\rho W_2-\pa_3 W_1=-g\eta_1\phi_2+g\eta_2\phi_1,
\end{aligned}
\label{BPS_Manton}
\end{align}
where $\epsilon = -\frac{1}{2}\mathrm{Re}\epsilon_1$ and we have chosen
the minus sign in the BPS equation.
These equations are invariant under
the gauge transformations \cite{FoHoPa}
\begin{eqnarray}
W'_1=W_1+{1\over g}\pa_\rho \Lambda,
&\quad&
W'_2=W_2+{1\over g}\pa_3 \Lambda, \nonumber \\
\phi'_1=\cos\Lambda \phi_1+\sin\Lambda \phi_2,
&\quad&
\phi'_2=\cos\Lambda \phi_2-\sin\Lambda \phi_1,\\
\eta'_1=\cos\Lambda \eta_1+\sin\Lambda \eta_2,
&\quad&
\eta'_2=\cos\Lambda \eta_2-\sin\Lambda \eta_1,\nonumber
\end{eqnarray}
where $\Lambda$ is a function of $(\rho,z)$.
There are five differential equations for the six unknown functions,
which have one gauge degree of freedom.
For $\epsilon=0$, the solution for
\eqref{BPS_Manton} has been found in
\cite{Prasad:1975kr,Bog}.
It is an interesting problem to find solutions to the deformed equation 
by perturbations around the exact solution 
and calculate the $\epsilon$-corrections
to the central charges. 
It would also be interesting to study the Nahm construction 
of monopoles \cite{Nahm:1979yw} and
other BPS solitons such as vortices
and domain walls.
These subjects will be discussed elsewhere.

\subsection*{Acknowledgment}
The work of S. S. is supported by the Japan Society
for the Promotion of Science (JSPS) Research Fellowship.

\end{document}